# Efficiency Theory:
## a Unifying Theory for Information, Computation and Intelligence

Roman V. Yampolskiy
Computer Engineering and Computer Science
University of Louisville, USA
roman.yampolskiy@louisville.edu

**Abstract**
The paper serves as the first contribution towards the development of the theory of efficiency: a unifying framework for the currently disjoint theories of information, complexity, communication and computation. Realizing the defining nature of the brute force approach in the fundamental concepts in all of the above mentioned fields, the paper suggests using efficiency or improvement over the brute force algorithm as a common unifying factor necessary for the creation of a unified theory of information manipulation. By defining such diverse terms as randomness, knowledge, intelligence and computability in terms of a common denominator we are able to bring together contributions from Shannon, Levin, Kolmogorov, Solomonoff, Chaitin, Yao and many others under a common umbrella of the efficiency theory.

**Introduction**
The quest for a Unified Theory of Everything (UTE) is well known to be a central goal in natural sciences. In recent years a similar aspiration to find a Unified Theory of Information (UTI) has been observed in computational sciences [1-13]. Despite numerous attempts, no such theory has been discovered and the quest to unify Shannon's Information Theory [14], Kolmogorov-Chaitin Complexity theory [15, 16], Solomonoff's Algorithmic Information Theory [17] and Yao's Communication complexity [18], as well as concepts of intelligence and knowledge continues. In this paper we present a novel set of definitions for information and computation related concepts and theories which is based on a common concept of efficiency. We show that a common thread exists and that future efforts could succeed in formalizing our intuitive notions. We further show some examples of how the proposed theory could be used to develop interesting variations on the current algorithms in communication and data compression.

**Efficiency Theory**
The proposed Efficiency Theory (EF) is derived with respect to the universal algorithm known as the "brute force" approach. Brute Force (BF) is an approach to solving difficult computational problems by considering every possible answer. BF is an extremely inefficient way of solving problems and is usually considered inapplicable in practice to instances of difficult problems of non-trivial size. It is an amazing and underappreciated fact that this simplest to discover, understand and implement algorithm also produces the most accurate (not approximate) solutions to the set of all difficult computational problems (NP-Hard, NP-Complete, etc.). In this paper we consider BF in an even broader context, namely, BF could be inefficient in other ways, for example representing otherwise compressible text strings by specifying every symbol.

Efficiency in general describes the extent to which resources such as time, space, energy, etc. are well used for the intended task or purpose. In complexity theory it is a property of algorithms for solving problems which require at most a number of steps (or memory locations) bounded from above by some polynomial function to be solved. The size of the problem instance is considered in determining the bounding function. Typically efficiency of an algorithm could be improved at the cost of solution quality. This often happens in cases where approximate solutions are acceptable. We also interpret efficiency to mean shorter representations of redundant data string. Essentially, EF measures how far can we get away from the BF in terms of finding quick algorithms for difficult problems studied in Complexity Theory (Levin [19], Cook [20], Karp [21], etc.) as well as towards discovering succinct string encodings (Shannon [14], Kolmogorov [15], Solomonoff [17], Chaitin [17]). Many fundamental notions related to information and computation could be naturally formalized in terms of their relevance to BF or efficiency.

**Information and Knowledge**
Information is a poorly understood concept and can be analyzed by different researchers from very different domain specific points of view [2]. Pervez assembled the following collection of definitions for the concept of information from over 20 different studies [22]:

- data that can be understood as a commodity or physical resource
- signal, code, symbol, message or medium
- formal or recovered knowledge
- subjective or personal knowledge
- thinking, cognition, and memory
- technology
- text
- uncertainty reduction
- linkage between living organisms and their environment
- product of social interaction that has a structure capable of changing the image structure of a recipient
- as a stimulus, information facilitates learning and acts as means for regulation and control in society

Hofkirchner [1] believes that the concept of information overlaps a number of concepts including: structure, data, signal, message, signification, meaning, sense, sign, sign process, semiosis, psyche, intelligence, perception, thought, language, knowledge, consciousness, mind, and wisdom.

Ever since Shannon presented his information theory, different approaches to measuring information have been suggested: Langefors' infological equation [23], Brookes' fundamental equation [24], Semantic Information Theory [25], and many others. In the proposed Efficiency Theory, *information* (Shannon [14], Hartley [26], Kelly [27]) measures how inefficiently knowledge (or specified information) is represented. (A special type of information sharing known as Communication Complexity [28] deals with the efficiency of communication between multiple computational processes and could be a subject to similar analysis). Shannon himself defined the fundamental problem of communication as that of "… reproducing at one point either exactly or approximately a message selected at another point. [14]" The BF approach to this problem would be to simply send over the whole message, symbol after symbol, completely disregarding any knowledge we might have about the properties of the text string in question. However, a more efficient approach may be to incorporate any knowledge we might already have about the message (for example that a certain symbol always starts any message) and to only transmit symbols which would reduce uncertainty about the message and by doing so provide us with novel knowledge.

From the above, ET allows us to define k*nowledge* as efficiently represented specified information. We are free to interpret the word efficiency either as an effective symbolic encoding or as an effective computation. Perhaps a few examples would help to define what we mean. With respect to efficient symbolic representation, Hoffman coding is a well-known example of an entropy encoding algorithm which uses variable-length codes calculated based on probability of occurrence of each source symbol to represent the message in the most efficient way [29]. The next example explains what we mean by efficient knowledge representation with respect to computation. If we want to share two numbers, we can do so in a number of ways. In particular, we can share the numbers in a direct and efficient, to retrieve, representation of knowledge: "3980750864240649373971255005503864911990643623425267084063851895759463889572617685 83317" and "4727721461074353025362230719730482246329146953020971164598521711305207112563635903 97527" or we can share the same two numbers, but in the form of necessary information, not efficiently accessible knowledge, as in, find the two factors of [30]:
"18819881292060796383869723946165043980716356337941738270076335642298885971523466548531906060 650474304531738801130339671619969232120573403187955065699622130516875930765025 7059". Both approaches encode exactly the same two numbers, only in the second case the recipient would have to spend a significant amount of computational resources (time) converting inefficiently presented data (information) into efficiently stored data (knowledge). Mizzaro suggests that the two types of information be referred to as "actual" and "potential" [2].

Another example aimed to illustrate information/knowledge distinction comes from an article by Aaronson [31]: The largest known prime number, as reported by the Mersenne.org, is p := $2^{43112609} - 1$. But what does it mean to say that p is "known"? Does that mean that, if we desired, we could print out all 30 pages of its decimal digits? That doesn't seem right. All that should really matter is that the expression '$2^{43112609} - 1$' picks out a unique positive integer, and that integer has been proven to be prime. However, if those are the criteria, then why can't we immediately beat the largest-known-prime record by postulating that: *p′ = The first prime larger than $2^{43112609} - 1$*. Clearly p′ exists and it is uniquely defined, and is also a prime number by definition. "If we want, we can even write a program that is guaranteed to find p′ and output its decimal digits, using a number of steps that can be upper-bounded a priori. Yet our intuition stubbornly insists that $2^{43112609} - 1$ is a "known" prime in a sense that p′ is not. Is there any principled basis for such a distinction? The clearest basis that I can suggest is the following. We know an algorithm that takes as input a positive integer k, and that outputs the decimal digits of p = $2^k - 1$ using a number of steps that is polynomial—indeed, linear—in the number of digits of p. But we do not know any similarly-efficient algorithm that provably outputs the first prime larger than $2^k - 1$ [31]."

Again, the only distinction between information and knowledge is how efficiently we can get access to the desired answer. In both cases we are dealing with pre-specified information since we know that the answer is going to represent a prime number, but knowledge is immediately available to us, while information may require an insurmountable amount of processing to deliver the same result. This leads us to an interesting observation: Information can't be created or destroyed, only made less efficiently accessible. For example prime numbers existed before the Big Bang and will continue to exist forever regardless of our best efforts to destroy them. At any point in time, one can simply start printing out a list of all integers and such a list will undoubtedly contain all prime numbers and as long as we are willing to extract specific numbers from such a list, our knowledge of particular prime numbers could be regained after paying some computational cost. Consequently that means that, knowledge could be created or destroyed by making it significantly less or more efficient to access or by providing or deleting associated specifications.

In fact we can generalize our prime number list example to the list of all possible strings of increasingly larger size. The idea of such a list is not novel and has been previously considered by Jorge Luis Borges in *The Library of Babel* [32], by Hans Moravec in *Robot* [33] and by Bruno Marchal in his PhD thesis [34]. Essentially, all the knowledge we will ever have is already available to us in the form of such string libraries. The only problem is that it is stored in an inefficient to access format, lacking specifications. The knowledge discovery process (computation) converts such inefficient information into easily accessible knowledge by providing descriptive pointers to optimally encoded strings to give them meaning. Specified information is a tuple (x,y) there f(x) has the same semantic meaning as y and function f is a specification. Given enough time we can compute any computable function so time is a necessary resource to obtain specified knowledge. Since multiple, in fact infinite, number of semantic pointers could refer to the same string [2] that means that a single string could contain an infinite amount of knowledge if taken in the proper semantic context, generating multiple levels of meaning. Essentially that means that obtained knowledge is relative to the receiver of information. It is mainly to avoid the resulting complications that Shannon has excluded semantics from his information theory [14].

Jurgen Schmidhuber has also considered the idea of string libraries and has gone so far as to develop algorithms for "Computing Everything" [35, 36]. In particular, concerned with the efficiency of his algorithm Schmidshuber has modified a Levin Search algorithm [37] to produce a provably fastest way to compute every string [35]. Schmidshuber's work shows that computing all information is easier than computing any specific piece, or in his words: "… computing all universes with all possible types of physical laws tends to be much cheaper in terms of information requirements than computing just one particular, arbitrarily chosen one, because there is an extremely short algorithm that systematically enumerates and runs all computable universes, while most individual universes have very long shortest descriptions [38]."

**Intelligence and Computation**
Computation is the process of obtaining efficiently represented information (knowledge) by any algorithm (including inefficient ones). Intelligence in the context of EF could be defined as the ability to design algorithms which are more efficient compared to brute force. Same ability shown for a variety of problems is known as general intelligence or universal intelligence [39].

An efficient algorithm could be said to exhibit intelligence in some narrow domain. In addressing specific instances of problems, an intelligent system can come up with a specific set of steps which don't constitute a general solution for all problems of such type, but are nonetheless efficient. Intelligence could also be defined as the process of obtaining knowledge by efficient means. If strict separation between different complexity classes (such as P VS NP) is proven, it would imply that no efficient algorithms for solving NP complete problems could be developed [40]. Consequently, that would imply that intelligence has an upper limit, a non-trivial result which has only been hinted at from limitations in physical laws and constructible hardware [41].

Historically, the complexity of computational processes has been measured either in terms of required steps (time) or in terms of required memory (space). Some attempts have been made in correlating the compressed (Kolmogorov) length of the algorithm with its complexity [42], but such attempts didn't find much practical use. We suggest that there is a relationship between how complex a computational algorithm is and intelligence, in terms of how much intelligence is required to either design or comprehend a particular algorithm. Furthermore we believe that such an intelligence based complexity measure is independent from those used in the field of complexity theory.

To illustrate the idea with examples we again will begin with the brute force algorithm. BF is the easiest algorithm to design as it requires very little intelligence to understand how it works. On the other hand an algorithm such as AKS primality test [43] is non-trivial to design or even to understand since it relies on a great deal of background knowledge. Essentially the intelligence based complexity of an algorithm is related to the minimum intelligence level required to design an algorithm or to understand it. This is a very important property in the field of education where only a certain subset of students will understand the more advanced material. We can speculate that a student with an "IQ" below a certain level can be shown to be incapable of understanding a particular algorithm. Likewise we can show that in order to solve a particular problem (P VS. NP) someone with IQ of at least X will be required. With respect to computational systems it would be inefficient to use extraneous intelligence resources to solve a problem for which a lower intelligence level is sufficient.

Consequently, efficiency is at the heart of algorithm design and so efficiency theory can be used to provide a novel measure of algorithm complexity based on necessary intellectual resources. Certain algorithms while desirable could be shown to be outside of human ability to design them because they are just too complex from the available intelligence-resources point of view. Perhaps the invention of superintelligent machines will make discovery/design of such algorithms feasible [44]. Also by sorting algorithms based on the perceived required IQ resources we might be able to predict the order in which algorithms will be discovered. Such an order of algorithm discovery would likely be consistent among multiple independently working scientific cultures, making it possible to make estimates of state-of-the-art in algorithm development. Such capability is particularly valuable in areas of research related to cryptography and integer factorization [45].

Given current state of the art in understanding of human and machine intelligence the proposed measure is not computable. However different proxy measures could be used to approximate the intellectual resources to solve a particular problem. For example the number of scientific papers published on the topic may serve as a quick heuristic to measure the problem's difficulty. Supposedly, in order to solve the problem one would have to be an expert in all of the relevant literature. As our understanding of human and machine intelligence increases a more direct correlation between available intellectual resources such as memory and difficulty of the problem will be derived.

**Time and Space**
In complexity theory time and space are the two fundamental measures of efficiency. For many algorithms time efficiency could be obtained at the cost of space and vice versa. This is known as space-time or time-memory tradeoff. With respect to communication memory size or the number of symbols to be exchanged in order to convey a message is a standard measure of communication efficiency. Alternatively, the minimum amount of time necessary to transmit a message can be used to measure the informational content of the message with respect to a specific information exchange system.

In the field of communication space-time efficiency tradeoffs could be particularly dramatic. It is interesting to look at two examples illustrating the extreme ends of the tradeoff spectrum appearing in synchronized communication [46]. With respect to the maximum space efficiency, communication with silence (precise measurement of delay) [47-51] represents the theoretical limit, as a channel with deterministic service time has infinite capacity [52]. In its simplest form in order to communicate with silence the sender transmits a single bit followed by a delay which if measured in pre-agreed upon units of time encodes the desired message [53]. The delay is followed by transmission of a second bit indicating termination of the delay. In real-life the communication system's network reliability issues prevent precise measurement of the delay, and consequently, transmission of arbitrarily large amount of information is impossible. However, theoretically silence based communication down to a Planck time is possible. Such a form of communication is capable of transmitting a large amount of information in a very short amount of time, approximately $10^{43}$ bits/s. Because precision of time communication could be detected, time itself could be used as a measure of communication complexity valid up to a multiplicative constant with respect to a particular communication system.

Alternatively, the same idea could be implemented in a way which uses computation instead of relying on access to a shared clock. Two sides wishing to communicate simultaneously start a program which acts as a simple counter and runs on identical hardware. Next, they calculate how long it takes to send a single bit over their communication channel (t). To send a message S the sender waits until S is about to be computed, and t time before that, sends 1 bit to the receiver who upon receiving the bit takes the counter value produced at that time as the message. At that point both parties start the cycle again. It is also possible and potentially more efficient with respect to time to cycle through all n-bit strings and by selecting appropriate n-bit segments construct the desired message. Such form of information exchange, once setup, essentially produced 1-bit communication which is optimally efficient from the point of view of required space. One bit communication is also energy efficient and may be particularly useful for interstellar communication with distinct satellites. This protocol is also subject to limitations inherent in the networking infrastructure and additional problems of synchronization.

**Compressibility and Randomness**
Kolmogorov Complexity (compressibility) is a degree of efficiency with which information could be represented. Information in its most efficient representation is essentially a random string of symbols. Correspondingly, degree of randomness is correlated to the efficiency with which information is presented. A string is algorithmically (Martin-Loef) random if it can't be compressed, or in other words its Kolmogorov complexity is equal to its length [54]. The Kolmogorov complexity of a string is incomputable, meaning that there is no efficient way of measuring it. Looking at the definition of knowledge presented in terms of efficiency theory we can conclude that randomness is pure knowledge. This is highly counterintuitive as outside of the field of information theory a random string of symbols is believed to contain no valuable patterns. However in the context of information theory randomness is a fundamental resource alongside time and space [55].

Compression paradox is an observation that a larger amount of information could be compressed more efficiently than a smaller more specified message. In fact taken to the extreme this idea shows that all possible information could be encoded in a program requiring just a few bytes as illustrated by Schmidhuber's algorithm for computing all universes [36, 38]. While two types of compression are typically recognized (lossy and lossless), compression paradox leads us to suggest a third variant we will call *Gainy Compression (GC)*.

GC works by providing a specification of original information to which some extra information if appended. GC keeps the quality of the message the same as original but instead reduces the confidence of the receiver that the message is in fact the intended message. Since in a majority of cases we are not interested in compressing random data, but rather files containing stories, movies, songs, passwords and other meaningful data, human intelligence can be used to separate semantically meaningful data from random noise. For example, an illegal prime is a number which if properly decoded represents information that is forbidden to possess or distribute [56]. One of the last fifty 100-million-digit-primes may happen to be an "illegal prime" representing a movie. Human intelligence can quickly determine which one just by looking at decoding of all fifty such primes in an agreed upon movie standard. So hypothetically, in some cases, we are able to encode a movie-segment with no quality loss in just a few bytes. This is accomplished by sacrificing time efficiency to gain space efficiency with the help of intelligence. Of course the proposed approach is itself subject to limitations of Kolmogorov complexity, particularly incommutability of optimal gainy compression with respect to decoding efficiency.

**Oracles and Undecidability**
Undecidability represents an absence of efficient or inefficient algorithms for solving a particular problem. A classic example of an undecidable problem is the halting problem, proven as such by Alan Turing [57]. Interestingly it was also Turing who suggested, what we will define as the logical complement to the idea of Undecidability, the idea of an Oracle [57]. With respect to efficiency theory we can define an oracle as an agent capable of solving a certain set of related problems with constant efficiency regardless of the size of the given problem instances. Some oracles are even capable of solving undecidable problems while remaining perfectly efficient. So an oracle for solving a halting problem can do so in a constant number of computational steps regardless of the size of the problem whose behavior it is trying to predict. In general oracles violate Rice's theorem with constant efficiency.

**Intractable and Tractable**
All computational problems could be separated into two classes: Intractable – a class of problems postulated to have no efficient algorithm to solve them and tractable – a class of efficiently solvable problems. The related P vs NP question which addresses possibility of finding efficient algorithms for intractable problems is one of the most important and well-studied problems of modernity [45, 58-65]. It is interesting to note that the number of tractable problems while theoretically infinite, with respect to those encountered in practice, is relatively small compared to the total number of problems in the mathematical universe, most of which are therefore only perfectly solvable by brute force methods [66].

**Conclusions and Future Directions**
All of the concepts defined above have a common factor, namely "efficiency," and could be mapped onto each other. First the constituent terms of pairs of opposites presented in Table 1 could be trivially defined as opposite ends of the same spectra. Next, some interesting observations could be made with respect to the relationships observed on less obviously related terms. For example, problems could be considered information, while answers to them are knowledge. Efficiency (or at least rudimentary efficient algorithms) could be produced by brute force approaches simply by trying all possible algorithms up to a certain length until a more efficient one is found. Finally, and somewhat surprisingly, perfect knowledge could be shown to be the same as perfect randomness. A universal efficiency measure could be constructed by contrasting the resulting solution with the pure brute force approach. So depending on the domain of analysis the ratio of symbols, computational steps, memory cells or communication bits to the number required by the brute force algorithm could be calculated as the normalized efficiency of the algorithm. Since the number of possible brute force algorithms is also infinite and they can greatly differ in their efficiency we can perform our analysis with respect to the most efficient brute force algorithm which works by considering all possible solutions, but not impossible ones.

Some problems in NP are solvable in practice, while some problems in P are not. For example an algorithm with running time of $1.00000001^n$ is preferred over the one with a running time of $n^{10000}$ [31]. This is a well-known issue and a limitation of a binary tractable/intractable separation of problems into classes. In our definition, efficiency is not a binary state but rather a degree ranging from perfectly inefficient (brute force required) to perfectly efficient (constant time solvable). Consequently the efficiency theory is designed to study the degree and limits of efficiency in all relevant domains of data processing.

Table 1: Base terms grouped in pairs of opposites with respect to efficiency.

| Efficient | Inefficient |
|---|---|
| Efficiency | Brute Force |
| Knowledge | Information |
| P | NP |
| Compressibility | Randomness |
| Intelligence | Computation |
| Space | Time |
| Oracle | Undecidable |

The proposed efficiency theory should be an important component of UTE and could have broad applications to fields outside of computer science, such as:

*Biology* Dead matter is inefficient, living matter is efficient in terms of obtaining resources, reproduction and problem solving. Proposed theory may be used to understand how via brute force trial and error living matter was generated from non-living molecules (a starting step for evolution and source of ongoing debate) [67].

*Education* We can greatly improve allocation of resources for education if we can calculate the most efficient level of intelligence required to learn any particular concept.

*Mathematics* Many subfields of mathematics have efficiency at their core. For example proofs of theorems require efficient verification [68]. Reductions between different problems used in complexity theory are also required to be more efficient compared to the computational requirements of the problems being reduced.

*Physics* The puzzling relationship between time and space in physics could be better understood via the common factor of computational efficiency. In fact many have suggested viewing the universe as a computational device [69-71].

*Theology* In most religions god is considered to be outside of the space-time continuum. As such god is not subject to issues of efficiency and may be interpreted as a Global Optimal Decider (GOD) for all types of difficult problems.

This paper serves as the first contribution to the development of the Efficiency Theory. In the future we plan on expanding EF to fully incorporate the following concepts which have efficiency as the core of their definitions:

- **Art, Beauty, Music, Novelty, Surprise, Interestingness, Attention, Curiosity, Science, Music, Jokes and Creativity** are byproducts of our desire to discover novel patterns by representing (compressing) data in efficient ways [72, 73].
- **Chaitin's incompleteness theorem** states that efficiency of a particular string can't be proven.
- **Computational irreducibility** states that other than running the software no more efficient way to predict behavior of a program (above a certain complexity level) exists [69].
- **Error correcting codes** are the most efficient way of correcting data transmission errors with fewest retransmissions.
- **Levin search** (Universal search) is a computable time (or space) bounded version of algorithmic complexity which measures efficiency of solving inversion problems [37].
- **Occam's razor** states that the most efficient (succinct) hypothesis fitting the data should be chosen over all others.
- **Paradoxes** are frequently based on violations of efficiency laws. For example Berry paradox: "the smallest possible integer not definable by a given number of words" is based on the impossibility of finding the most efficient representation for a number.

- **Potent numbers** proposed by Adleman are related to the Kolmogorov and Levin complexity and take into account the amount of time required to generate the string in question in the most efficient way [55].
- **Pseudorandomness** in computational complexity is defined as a distribution which can't be efficiently distinguished from the uniform distribution.
- **Public key cryptography** is perfectly readable without a key but not efficiently (will take millions of years to read a message with current software/hardware).
- **Recursive self-improvement** in software continuously improves efficiency of resource consumption and computational complexity of intelligent software [74-76].

## References

[1] W. Hofkirchner, "How To Achieve a Unified Theory of Information," *Triple C - Cognition, Communication, Co-operation,* vol. 7(2), pp. 357-368, 2009.
[2] S. Mizzaro, "Towards a Theory of Epistemic Information," *Information Modelling and Knowledge Bases,* vol. 12, pp. 1-20, 2001.
[3] J. Holmstrom and T. Koli, "Making the Concept of Information Operational," *Department of Information Technology and Media,* vol. Mid Sweden University, pp. Master Thesis in Informatics, http://www.palmius.com/joel/lic/infoop.pdf, 2002.
[4] M. Burgin, "Information: Paradoxes, Contradictions, and Solutions," *Triple C - Cognition, Communication, Co-operation,* vol. 1(1), pp. 53-70, 2003.
[5] L. Floridi, "What Is The Philosophy of Information?," *Metaphilosophy,* vol. 33(1-2), pp. 123-145, 2002.
[6] W. Hofkirchner, "A Unified Theory of Information As Transdisciplinary Framework," presented at the ICT&S Center for Advanced Studies and Research, http://www.idt.mdh.se/ECAP-2005/articles/BIOSEMANTICS/WolfgangHofkirchner/WolfgangHofkirchner.pdf, 2005.
[7] W. Hofkirchner, "Cognitive Sciences In the Perspective of a Unified Theory of Information," presented at the 43rd Annual Meeting of the International Society for the Systems Sciences (ISSS), Pacific Grove, CA, 1999.
[8] S. Ji, "Computing With Numbers, Words, or Molecules: Integrating Mathematics, Linguistics and Biology Through Semiotics," in *Reports of the Research Group on Mathematical Linguistics*, Taqrragona, Spain, 2003, pp. 1-12.
[9] G. M. Braga, "Semantic Theories of Information," *Information Sciences (Ciência da Informação),* vol. 6(2), pp. 69-73, 1977 1977.
[10] Y. X. Zhong, "Mechanism Approach to a Unified Theory of Artificial Intelligence," presented at the IEEE International Conference on Granular Computing, Beijing, 2005.
[11] F. Fluckiger, "Towards a Unified Concept of Information: Presentation of a New Approach," *World Futures: The Journal of General Evolution,* vol. 49/50 (3-4), p. 309, July 1997.
[12] D. Doucette, R. Bichler, W. Hofkirchner, and C. Raffl, "Toward a New Science of Information," *Data Sciences Journal,* vol. 6(7), pp. 198-205, 2007.
[13] P. Fleissner and W. Hofkirchner, "Emergent Information: Towards a Unified Information Theory," *BioSystems,* vol. 38(2-3), pp. 243-248, 1996.
[14] C. E. Shannon, "A Mathematical Theory of Communication," *Bell Systems Technical Journal,* vol. 27(3), pp. 379-423, July 1948.
[15] A. N. Kolmogorov, "Three Approaches to the Quantitative Definition of Information," *Problems Inform. Transmission,* vol. 1(1), pp. 1-7, 1965.

[16]    G. J. Chaitin, "On the Length of Programs for Computing Finite Binary Sequences," *Journal of the ACM (JACM),* vol. 13(4), pp. 547-569, 1966.
[17]    R. Solomonoff, "A Preliminary Report on a General Theory of Inductive Inference " presented at the Report V-131, Zator Co., Cambridge, Ma., February 4, 1960.
[18]    A. C. Yao, "Some Complexity Questions Related to Distributed Computing," presented at the 11th Symposium on Theory of Computing (STOC), 1979.
[19]    L. Levin, "Average-case complete problems," *SIAM J. Comput,* vol. 15, pp. 285-286, 1986.
[20]    S. A. Cook and R. A. Reckhow, "The Relative Efficiency of Propositional Proof Systems," *The Journal of Symbolic Logic,* vol. 44(1), pp. 36-50, 1979.
[21]    R. M. Karp, "Reducibility Among Combinatorial Problems," in *Complexity of Computer Computations*, R. E. Miller and J. W. Thatcher, Eds., ed New York: Plenum, 1972, pp. 85-103.
[22]    A. Pervez, "Information As Form," *Triple C - Cognition, Communication, Co-operation,* vol. 7(1), pp. 1-11, 2009.
[23]    B. Langefors, *Theoretical analysis of information systems*. Lund: Studentlitteratur, 1966.
[24]    B. C. Brookes, "The fundamental equation of information science," *Problems of Information Science,* vol. 530, pp. 115-130, 1975.
[25]    J. Hintikka, *On Semantic Information. In: Physics, Logic, and History*: Plenum Press, 1970.
[26]    R. V. L. Hartley, "Transmission of Information " *Bell System Technical Journal* vol. 7(3), pp. 535-563, July 1928.
[27]    J. L. Kelly, "A New Interpretation of Information Rate," *Bell System Technical Journal,* vol. 35, pp. 917-926, 1956.
[28]    A. C. Yao, "Some Complexity Questions Related to Distributed Computing," presented at the Proc. of 11th STOC, 1979.
[29]    D. A. Huffman, "A Method for the Construction of Minimum-Redundancy Codes," presented at the Institute of Radio Engineers (I.R.E.), September 1952.
[30]    (2007). *The RSA Challenge Numbers*.
[31]    S. Aaronson, "Why Philosophers Should Care About Computational Complexity," in *Computability: Godel, Turing, Church, and beyond*, ed: MIT Press, 2012.
[32]    J. L. Borges, *The Library of Babel* Jeffrey, New Hemshire: David R. Godine Publisher 2000.
[33]    H. Moravec, *Robot*: Wiley Interscience, 1999.
[34]    B. Marchal, "Calculabilite, Physique et Cognition," presented at the PhD thesis, L'Universite des Sciences et Technologies De Lilles, 1998.
[35]    J. Schmidhuber, "The Speed Prior: A New Simplicity Measure Yielding Near-Optimal Computable Predictions," presented at the 15th Annual Conference on Computational Learning Theory (COLT 2002), Sydney, Australia, 2002.
[36]    J. Schmidhuber, "A Computer Scientist's View of Life, the Universe, and Everything," *LNCS, Springer,* pp. 201-288, 1997.
[37]    L. Levin, "Universal Search Problems," *Problems of Information Transmission,* vol. 9(3), pp. 265--266, 1973.
[38]    J. Schmidhuber, "Algorithmic Theories of Everything," Available at: http://arxiv.org/pdf/quant-ph/0011122v2, November 30, 2000.
[39]    S. Legg and M. Hutter, "Universal Intelligence: A Definition of Machine Intelligence," *Minds and Machines,* vol. 17(4), pp. 391-444, December 2007.
[40]    R. V. Yampolskiy, "Construction of an NP Problem with an Exponential Lower Bound," *Arxiv preprint arXiv:1111.0305,* 2011.
[41]    S. Lloyd, "Ultimate Physical Limits to Computation," *Nature,* vol. 406, pp. 1047-1054, 2000.
[42]    B. A. Trakhtenbrot, "A Survey of Russian Approaches to Perebor (Brute-Force Searches) Algorithms," *IEEE Annals of the History of Computing,* vol. 6(4), pp. 384-400, 1984.

[43]  M. Agrawal, N. Kayal, and N. Saxena, "PRIMES is in P," *Annals of Mathematics,* vol. 160(2), pp. 781-793, 2004.
[44]  R. V. Yampolskiy, "AI-Complete CAPTCHAs as Zero Knowledge Proofs of Access to an Artificially Intelligent System," *ISRN Artificial Intelligence,* vol. 2012(271878), 2011.
[45]  R. V. Yampolskiy, "Application of Bio-Inspired Algorithm to the Problem of Integer Factorisation," *International Journal of Bio-Inspired Computation (IJBIC),* vol. 2(2), pp. 115-123, 2010.
[46]  R. Impagliazzo and R. Williams, "Communication complexity with synchronized clocks," presented at the IEEE 25th Annual Conference on Computational Complexity (CCC '10), Washington, DC, USA, 2010.
[47]  C. Fragouli and A. Orlitsky, "Silence is Golden and Time is Money: Power-Aware Communication for Sensor Networks," presented at the 43rd Allerton conference, 2005.
[48]  A. K. Dhulipala, C. Fragouli, and A. Orlitsky, "Silence-Based Communication," *IEEE Transactions on Information Theory,* vol. 56(1), pp. 350-366, 2010.
[49]  J. Giles and B. Hajek, "An information-theoretic and game-theoretic study of timing channels," *IEEE Transactions on Information Theory,* vol. 48(9), pp. 2455-2477, 2002.
[50]  R. Sundaresan and S. Verdu, "Robust decoding for timing channels," *IEEE Transactions on Information Theory,* vol. 46(2), pp. 405-419, 2000.
[51]  A. S. Bedekar and M. Azizoglu, "The information-theoretic capacity of discrete-time queues," *IEEE Transactions on Information Theory* vol. 44(2), pp. 446-461, 1998.
[52]  V. Anantharam and S. Verdu, "Bits through queues," *IEEE Transactions on Information Theory,* vol. 42(1), pp. 4-18, 1996.
[53]  S. Cabuk, C. E. Brodley, and C. Shields, "IP covert timing channels: design and detection," presented at the 11th ACM conference on Computer and communications security, Washington DC, USA, 2004.
[54]  P. Martin-Löf, "The definition of random sequences," *Information and Control,* vol. 9(6), pp. 602-619, 1966.
[55]  L. Adleman, "Time, Space, and Randomness," *Technical Report: MIT/LCS/TM-131,* April 1979.
[56]  Anonymous, "Illegal Prime," presented at the WIkipedia, Available at: http://en.wikipedia.org/wiki/Illegal_prime, Retrieved December 6, 2011.
[57]  A. Turing, "On computable numbers, with an application to the Entscheidungsproblem," *Proceedings of the London Mathematical Society,* vol. 2(42), pp. 230-265, 1936.
[58]  S. Aaronson, "Why Philosophers Should Care About Computational Complexity," *http://www.scottaaronson.com/papers/philos.pdf,* August 2011.
[59]  S. Aaronson, "NP-Complete Problems and Physical Reality," *ACM SIGACT News,* vol. 36(1), pp. 30-52, March 2005.
[60]  S. Aaronson, "Is P versus NP formally independent?," *Bulletin of the European Association for Theoretical Computer Science,* vol. 81, pp. 109-136, October, 2003.
[61]  S. Cook, "The P Versus NP Problem," *http://www.claymath.org/millennium/P_vs_NP/Official_Problem_Description.pdf,* pp. 1-19.
[62]  C. H. Papadimitriou, "NP-Completeness: A Retrospective," presented at the 24th International Colloquium on Automata, Languages and Programming (ICALP '97), Bologna, 1997.
[63]  B. A. Trakhtenbrot, "A Survey of Russian Approaches to Perebor (Brute-Force Search) Algorithms," *Annals of the History of Computing,* vol. 6(4), pp. 384-397, October 1984.
[64]  A. Wigderson, "P, NP and Mathematics - A Computational Complexity Perspective," presented at the International Conference of Mathematicians (ICM '06), Madrid, 2006.

[65]	R. V. Yampolskiy and A. EL-Barkouky, "Wisdom of Artificial Crowds Algorithm for Solving NP-Hard Problems," *International Journal of Bio-Inspired Computation (IJBIC),* vol. 3(6), pp. 358-369, 2011.
[66]	M. R. Garey and D. S. Johnson, *Computers and Intractability: A Guide to the Theory of NP-Completeness*: W. H. Freeman and Company, 1979.
[67]	R. Dawkins, *The Selfish Gene*. New York City: Oxford University Press, 1976.
[68]	A. Wigderson, "Knowledge, Creativity and P versus NP," Available at: http://www.math.ias.edu/~avi/PUBLICATIONS/MYPAPERS/AW09/AW09.pdf, 2009.
[69]	S. Wolfram, *A New Kind of Science*: Wolfram Media, Inc, May 14, 2002.
[70]	K. Zuse, *Rechnender Raum*. Braunschweig: Friedrich Vieweg & Sohn, 1969.
[71]	E. Fredkin, "Finite Nature," presented at the XXVIIth Rencotre de Moriond Series: Moriond Workshops, Les Arcs, Savoie, France, January 25 - February 1, 1992.
[72]	J. Schmidhuber, "Formal Theory of Creativity, Fun, and Intrinsic Motivation (1990-2010)," *IEEE Transactions on Autonomous Mental Development,* vol. 2(3), pp. 230-247, 2010.
[73]	J. Schmidhuber, "Simple Algorithmic Theory of Subjective Beauty, Novelty, Surprise, Interestingness, Attention, Curiosity, Creativity, Art, Science, Music, Jokes," *Journal of SICE,* vol. 48(1), pp. 21-32, 2009.
[74]	S. M. Omohundro, "The Nature of Self-Improving Artificial Intelligence," presented at the Singularity Summit, San Francisco, CA, 2007.
[75]	J. S. Hall, "Self-Improving AI: An Analysis," *Minds and Machines,* vol. 17(3), pp. 249 - 259, October 2007.
[76]	R. V. Yampolskiy, L. Reznik, M. Adams, J. Harlow, and D. Novikov, "Resource awareness in computational intelligence," *International Journal of Advanced Intelligence Paradigms,* vol. 3, pp. 305-322, 2011.